\documentclass[convention,peer-reviewed]{aesconf}

\graphicspath{{./}{figures/}}

\usepackage[utf8]{inputenc}

\usepackage{microtype}

\usepackage[numbers,square]{natbib}

\usepackage{booktabs}
\usepackage{color}
\usepackage{url}
\usepackage{amsmath}

\makeatletter
\def\@@eqno{\pdfprimitive\eqno}
\def\@@leqno{\pdfprimitive\leqno}

\makeatother

\usepackage{amssymb}
\usepackage{calrsfs}
\usepackage[table,xcdraw]{xcolor}
\usepackage{gensymb}
\DeclareMathAlphabet{\pazocal}{OMS}{zplm}{m}{n}

\newcommand{\SI}[2]{#1\,#2}
\newcommand{\hertz}{Hz}
\newcommand{\kilo}{k}
\newcommand{\milli}{m}

\newcommand{\second}{s}
\newcommand{\decibel}{dB}



\title{A Survey of Methods for the Discretization of Phonograph Record Playback Filters}

\author[1]{Benjamin R. Thompson}
\author[1]{Tre DiPassio}
\author[1]{Jenna Rutowski}
\author[1]{Michael C. Heilemann}

\affil[1]{University of Rochester}

\correspondence{Benjamin R. Thompson}{benjaminrthompson.mail@gmail.com}

\lastnames{Thompson, DiPassio, Rutowski, and Heilemann}

\shorttitle{Discretization of Phonograph Record Playback Filters}

\begin{document}

\twocolumn[
\maketitle 

\begin{onecolabstract}
Since the inception of electrical recording for phonograph records in 1924, records have been intentionally cut with a non-uniform frequency response to maximize the information density on a disc and to improve the signal-to-noise ratio. To reproduce a nominally flat signal within the available bandwidth, the effects of this cutting curve must be undone by applying an inverse curve on playback. Until 1953, with the introduction of what has become known as the RIAA curve, the playback curve required for any particular disc could vary by record company and over time. As a consequence, anyone seeking to hear or restore the information on a disc must have access to equipment that is capable of implementing multiple playback equalizations. This correction may be accomplished with either analog hardware or digital processing. The digital approach has the advantages of reduced cost and expanded versatility, but requires a transformation from continuous time, where the original curves are defined, to discrete time. This transformation inevitably comes with some deviations from the continuous-time response near the Nyquist frequency. There are many established methods for discretizing continuous-time filters, and these vary in performance, computational cost, and inherent latency. In this work, several methods for performing this transformation are explored in the context of phonograph playback equalization, and the performance of each approach is quantified. This work is intended as a resource for anyone developing systems for digital playback equalization or similar applications that require approximating the response of a continuous-time filter digitally.\end{onecolabstract}
]

\section{Introduction}

Almost all electrically recorded records are cut with an equalization curve that allows the cutter to switch between a constant-velocity response and a constant-amplitude response for different bands within the audible spectrum. For laterally-cut grooves, this allows for smaller groove widths, and thus more information over a given surface area. When high-frequency pre-emphasis is included in the cutting profile, it has the additional effect of increasing the overall signal-to-noise ratio upon playback as high-frequency surface noise is suppressed. When a record is played, an equalization that is the inverse of the cutting curve must be applied to accurately decode the information on the disc \cite{galo1996disc,recordingRepro}. 

From 1924, when the research arm of Western Electric, Bell Labs, introduced a system for electrically recording phonograph records~\cite{maxfield1926high}, until the 1950s when the recording industry settled on the RIAA curve as the standard \cite{RIAAstand}, the equalization applied when cutting a master recording varied from record company to record company, and over time as the technology evolved. The properties of some of these curves, particularly for smaller record companies, are unpublished, and even for the most widely used curves, there is disagreement among professionals about their exact properties~\cite{galo2009columbia,leister2017london}. Additionally, even when cutting equalization information is known for a particular label, it is often unclear which curve was utilized for any individual release~\cite{copeland2008manual}. As a result, a restoration engineer working with disc recordings must have the ability to implement multiple established playback curves, as well as the means to adjust the properties of the equalization \textit{in situ} to suit a particular recording.

This equalization is often performed using analog equipment. Because all playback curves were originally defined in continuous time, if the cutting transfer function is known, an inverse transfer function can be implemented more or less exactly with this approach. However, many analog playback equalizers are limited to a relatively small number of turnover frequencies for each filter element \cite{VADLYD, McIntosh}, with the consequence that, if a required turnover frequency falls outside of those values, the correct equalization cannot be implemented with that device. In addition, when performing playback equalization using an analog equalizer in the context of digitizing a disc, it is common to apply the playback curve before recording. If, at some later time, it is decided that a different curve is desirable, the information must be re-recorded. For historic recordings, this may be unacceptable because of the degradation each playback imposes on the medium. Instead, if the signal is recorded without equalization and the equalization is applied in the digital domain, it can be adjusted without consequence to the original media. Digital approaches are more versatile and less expensive than their analog counterparts, however, it is not possible to exactly replicate the response of an analog filter at all frequencies with a digital filter.  

The following sections present a survey of several established methods for discretizing continuous-time filters in the context of implementing phonograph playback equalization in the digital domain.

\section{The RIAA Curve as a Prototype Continuous-Time Filter}\label{sect:RIAA}

What is now known as the RIAA curve was developed by RCA Victor as the "New Orthophonic" curve and was published in 1953 \cite{moyer1953evolution}. This curve was adopted as a standard by a number of recording and broadcast organizations including the Recording Industry Association of America, eventually becoming a universal standard applied to all vinyl records \cite{copeland2008manual}. 

The RIAA playback curve makes an appropriate platform for evaluating the methods presented here because it contains high-frequency de-emphasis resulting in the presence of features near what will become the Nyquist frequency, which makes it challenging to discretize accurately. Additionally, its ubiquity means that it would likely be included in any system for applying playback equalization digitally.

The original RIAA playback curve is defined by three time constants: $\tau_0 = 318\mu s$, $\tau_1 = 75\mu s$, and $\tau_2 = 3180\mu s$. Though an amendment adding a fourth time constant at $7950\mu s$ was adopted in 1976, it was later withdrawn~\cite{self2020small}, and in this article, we will use only the original three time constants. The transfer function for this filter, with a gain factor applied such that the magnitude response at \SI{1}{\kilo\hertz} is approximately equal to \SI{0}{\decibel}, is  
\begin{equation}\label{eq:Hpz}   
    H(s)=\frac{\frac{1}{\tau_1}\left(s + \frac{1}{\tau_0}\right)}{\left(s + \frac{1}{\tau_1}\right)\left(s + \frac{1}{\tau_2}\right)}.
\end{equation}
Expressed as a ratio of polynomials, the transfer function becomes, 
\begin{equation}\label{eq:H}   
    H(s)=\frac{\frac{1}{\tau_1}s+\frac{1}{\tau_1 \tau_0}}{s^{2}+\frac{\tau_1+\tau_2}{\tau_1\tau_2}s+\frac{1}{\tau_1\tau_2}}.
\end{equation}

\section{Discretization Methods}\label{sect:methods}

Eight discretization methods were chosen to be evaluated in the context of playback equalization. Each of the methods is order-preserving in the sense that the order of the denominator of the transfer function, $N$, remains unchanged by the discretization. However, if the order of the numerator of the continuous-time filter, $M_s$, is less than $N$, it may be increased by the discretization process such that the order of the numerator of the discrete-time filter, $M_z$, becomes equal to $N$. The following subsections contain a brief theoretical overview of each method.

\subsection{Zero-Order Hold}\label{sec:ZOH}

This method entails approximating the output of a continuous filter, $H(s)$, when the input signal, $x[n] = x(nT)$, is a sampled representation of a continuous-time signal, $x(t)$, with a sampling period of $T$ and integer sample index $n$. For a zero-order hold, the sample value at $x[n]$ is directly held constant for the entire sampling period, resulting in the $x(t) = x(nT) = x[n]$ for the interval from $nT \leq t < (T + nT)$. This results in an approximated continuous-time signal that resembles a piece-wise function. From \cite{franklin1998digital}, the discrete-time transfer function, $H(z)$, may be approximated from $H(s)$ as,
\begin{equation}\label{eq:ZOH}  
    H(z) = \left(1-z^{-1}\right)\pazocal{Z}\left[\frac{H(s)}{s}\right],
\end{equation}
\noindent where $\pazocal{Z}$ is the Z-transform. In this work, the Zero-Order Hold method was implemented using MATLAB's \textit{c2d} function \cite{MATLABc2d}, with the Zero Order Hold (`zoh') method \cite{MATLABmethods} selected.

\subsection{Triangle Approximation}

This hold-based method is also referred to as a first-order hold and is similar to the Zero-Order Hold method discussed in Sect.~\ref{sec:ZOH}. For the Triangle Approximation, the continuous signal, $x(t)$, is approximated in the sampling interval from $nT \leq t < (T + nT)$ by connecting adjacent sampled values $x(nT)$ and $x(T+nT)$ with a straight line. This method is related to linear interpolation, which is not a causal process in continuous time. However, following \cite{franklin1998digital}, the discrete-time equivalent process is causal, and is given by,
\begin{equation}\label{eq:FOH}   
    H(z) = \frac{ \left(1-z^{-1}\right)^2}{Tz}\pazocal{Z}\left[\frac{H(s)}{s^2}\right].
\end{equation}
\noindent In this work, the Triangle Approximation method was implemented using MATLAB's \textit{c2d} function \cite{MATLABc2d}, with the First-Order Hold ('foh') method \cite{MATLABmethods} selected.

\subsection{Impulse Invariant Method}
The Impulse Invariant method, as outlined in \cite{smith2010physical}, produces a discrete-time transfer function whose impulse response is identical to the continuous-time impulse response at intervals of $nT$. The frequency response of the $z$-domain transfer function, however, is an aliased version of the frequency response of the $s$-domain transfer function. 

To perform a transformation using the Impulse Invariant method, a partial fraction expansion (PFE) to first-order terms is performed to find the residues (numerators of first-order fractions after PFE) and poles of the continuous-time system. For the prototype $s$-domain filter specified by Eq.~\ref{eq:H}, the transfer function after the PFE takes the form,

\begin{equation}\label{Hpfe}   
    H(s)= \frac{\frac{\tau_1\tau_2-\tau_0\tau_2}{\tau_0\tau_1^{2}-\tau_0\tau_1\tau_2}}{s+\frac{1}{\tau_1}}+\frac{\frac{\frac{\tau_2}{\tau_0}-1}{\tau_2-\tau_1}}{s+\frac{1}{\tau_2}},
\end{equation}

where $\tau_0$, $\tau_1$, and $\tau_2$ are the time constants defined in Sect. \ref{sect:RIAA}.

The poles of the digital filter are defined by transforming the poles of the analog filter with the substitution, 
\begin{equation}\label{eq:zpmatch}
    z=e^{sT}.
\end{equation}
 The residues of the digital filter are defined to be the same as the residues of the continuous-time system scaled by a factor of $T$. The digital transfer function for this case is given by,
\begin{equation}\label{HImpInv}   
    \hat{H}(z) = \frac{T\frac{\tau_1\tau_2-\tau_0\tau_2}{\tau_0\tau_1^{2}-\tau_0\tau_1\tau_2}}{z-e^{\frac{-1}{\tau_1}T}}+\frac{T\frac{\frac{\tau_2}{\tau_0}-1}{\tau_2-\tau_1}}{z-e^{\frac{-1}{\tau_2}T}}.
\end{equation}
The first-order terms defined by these new poles and residues can then be added together to generate a new ratio of polynomials in the $z$-domain.  

In this work, the Impulse Invariant method was implemented using MATLAB's \textit{c2d} function \cite{MATLABc2d}, with the Impulse-Invariant Mapping (`impulse') method selected \cite{MATLABmethods}. 

\subsection{Bilinear Transform}\label{sec:bilinear}

In the bilinear transform, or Tustin approximation, the $j\omega$ axis in the $s$-plane is mapped to the unit circle in the $z$-plane. Because the $j\omega$ axis is infinite, and the unit circle is finite, this results in some frequency warping, meaning that behavior at one frequency in the continuous-time system may be shifted to a different frequency in the discrete-time system. This frequency warping is most pronounced near the Nyquist frequency. There is a modification to the bilinear transform known as pre-warping that allows perfect matching between the continuous and discrete-time filter responses at one frequency. This can be useful if the filter has a significant feature at one frequency, such as the center frequency of a band-pass filter. In this work, the bilinear transform for the application of playback equalization is considered without the inclusion of pre-warping, though a method of pre-mapping the continuous time spectrum to minimize the frequency warping incurred by performing the bilinear transform is explored in Sect.~\ref{sect:Nyquist}.

The bilinear transform from the $s$-domain to the~$z$-domain is accomplished by making the substitution,
\begin{equation}\label{eq:sqError}   
    s = \frac{2}{T}\frac{z-1}{z+1}.
\end{equation}

For the purposes of this work, the Bilinear Transform method was implemented using MATLAB's \textit{c2d} function \cite{MATLABc2d}, with the Tustin (`tustin') method selected \cite{MATLABmethods}.

\subsection{Zero-Pole Matching}

Zero-Pole Matching (also called the Matched-$z$ Transformation \cite{smith2010physical}) is accomplished by mapping $s$-domain poles and zeros to poles and zeros in the $z$-domain directly. Each pole from the continuous system is mapped to a pole in the discrete system by the equivalency in Eq.~\ref{eq:zpmatch} \cite{franklin1998digital}. The $s$-domain zeros are also mapped to the $z$-domain by Eq. \ref{eq:zpmatch}, with the exception of $s=\infty$, which is ignored in this analysis as it is not relevant to this application.

For the prototype filter, combining Eq.~\ref{eq:zpmatch} with Eq.~\ref{eq:Hpz} gives,
\begin{equation}\label{Hpz}   
    \hat{H_0}(z)=\frac{\left(z - e^{\frac{-1}{\tau_0}T}\right)}{\left(z - e^{\frac{-1}{\tau_1}T}\right)\left(z - e^{\frac{-1}{\tau_2}T}\right)}.
\end{equation}
The gain of the digital filter is then set to match the gain of the analog filter in Eq.~\ref{eq:Hpz} at \SI{0}{\hertz}, resulting in a discretized transfer function,
\begin{equation}\label{Hhat}   
    \hat{H}(z)=G\hat{H_0}(z),
\end{equation}
where
\begin{equation}\label{G}   
    G=\frac{H(s=0)}{\hat{H_0}(z=1)}.
\end{equation}

In this work, the Zero-Pole Matching method was implemented using MATLAB's \textit{c2d} function \cite{MATLABc2d}, with the Zero-Pole Matching Equivalents method (`matched') selected \cite{MATLABmethods}.

\subsection{Complex Error Minimization} \label{sect:itCompErrMin}

This method determines the numerator and denominator of the digital filter, $\hat{B}(z)$ and $\hat{A}(z)$ respectively, by minimizing the weighted squared error between the complex frequency response of the continuous filter, $H(j\omega)$, and the complex frequency response of the discrete filter, $\hat{H}(j\omega)$, given by
\begin{equation}
    \frac{\hat{B}(j\omega)}{\hat{A}(j\omega)}.
\end{equation}
This is accomplished in two stages. First, the weighted squared equation error is minimized. For a desired sample rate of $f_{s}$, this minimization takes the form,
\begin{equation}\label{eq:eqError}   
     \min_{\hat{B}_0,\hat{A}_0} \sum_{k=0}^{n-1}{w_k|\hat{A}(j\omega_k)H(j\omega_k)-\hat{B}(j\omega_k)|^2},
\end{equation}
where $\hat{B}_0$ and $\hat{A}_0$ are the numerator and denominator of the digital filter at this stage respectively, $\omega$ is a length-$n$ vector of equally-spaced angular frequencies spanning from $\omega = 0$ rad/s to $\omega = \frac{n-1}{n}\pi f_{s}$ rad/s, $w$ is a length-$n$ array of weights, and $k$ is an indexing variable. The minimization in Eq.~\ref{eq:eqError} can be accomplished in a single iteration by solving a system of linear equations. 

In the second stage of the algorithm, the results of the equation error minimization, $\hat{B}_0$ and $\hat{A}_0$, are used as the initial estimate for an iterative minimization of the weighted squared output error using the damped Gauss-Newton method. This minimization takes the form,  
\begin{equation}\label{eq:sqError}   
     \min_{\hat{B}(z),\hat{A}(z)} \sum_{k=0}^{n-1}{w_k|H(j\omega_k)-\hat{H}(j\omega_k)|^2}.
\end{equation}
For the purposes of this work, the Complex Error Minimization method was implemented using MATLAB's \textit{invfreqz} function \cite{MATLABinvFreq}. The maximum number of iterations was set to 100 and the gradient vector norm threshold for convergence was set to 0.001.

\subsection{Magnitude Error Minimization}

This method is similar to the Complex Error Minimization method with the exception that it takes in as its argument only the sampled magnitude response of the prototype analog filter, and then constructs a minimum-phase complex response around it using the cepstral method \cite{smith2007introduction}. This is then fed into the Complex Error Minimization algorithm described in Sect.~\ref{sect:itCompErrMin}. Compared to the Complex Error Minimization method, this method yields a discrete filter with a magnitude response closer to that of the prototype filter, at the expense of a larger error in the phase response.


\subsection{Nyquist Band Transform}
\label{sect:Nyquist}
The Nyquist Band Transform (NBT)~\cite{darabundit2022nyquist} uses conformal mapping to pre-map a continuous-time filter so that when it is discretized using the bilinear transform, the amount of frequency warping is significantly reduced when compared to using the bilinear transform alone.

As discussed in Sect.~\ref{sec:bilinear}, the bilinear transform results in frequency warping because it compresses the infinite range of frequencies along the positive $j\omega$ axis in the $s$-domain, $[0,\infty)$ radians per second, to the finite range of frequencies along the positive half of the unit circle in the $z$-domain, $[0,\pi)$ radians per sample. The NBT attempts to address this by pre-mapping the Nyquist band in the $s$-domain so that $\omega = [0,\omega_0)$ maps to $\omega~=~[0,\infty)$, where $\omega_0$ is one half of the desired sample rate in radians per seconds. The ideal result is that, after performing the bilinear transform, the Nyquist band in the $s$-domain maps to the Nyquist band in the $z$-domain.

A detailed description of the NBT algorithm can be found in Appendix~\ref{NBT}.

\section{Oversampling}

The filters produced by each of the discretization methods presented in this work suffer from some amount of error when compared to the continuous-time prototype filter. In the frequency domain, this error is most pronounced near the Nyquist frequency. This is particularly an issue for audio signals recorded at sample rates where the Nyquist frequency is near the upper limit of the audible band. One common method of mitigating this is to move the Nyquist frequency further away from the audible band by oversampling. In this approach, the audio signal to be filtered is first upsampled, then filtering is performed at this higher sampling rate, after which the signal is downsampled back to the original sample rate. 

\begin{figure}[!t] 
\begin{center}
\includegraphics[width=0.95\columnwidth]{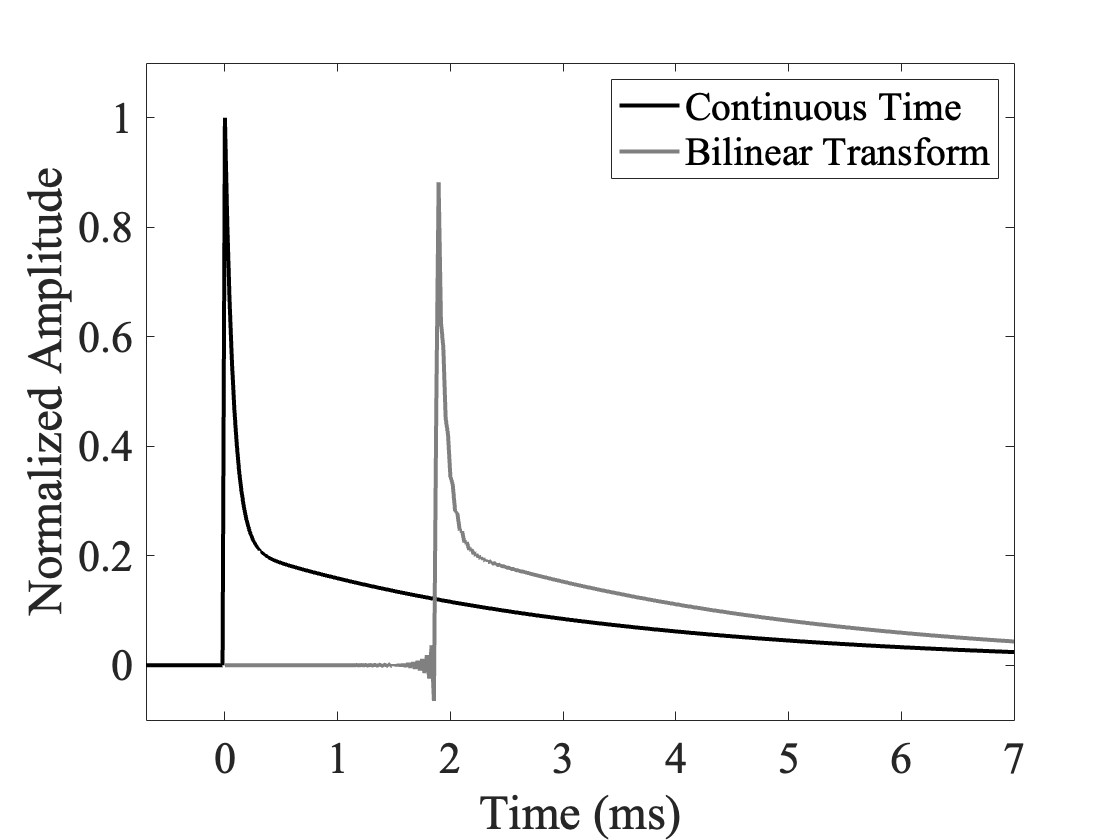}
\caption{The sampled impulse response of the continuous-time prototype filter along with the impulse response of a discretized version generated by the bilinear transform and an oversampling factor of $2$. The anti-aliasing filters used in this work result in a delay of \SI{1.87}{\milli\second} for this configuration. The pre-ringing introduced by the oversampling process can be observed in the discretized filter impulse response just before the \SI{2}{\milli\second} mark.}
\label{fig:ovrSamp}
\end{center}
\end{figure}

Oversampling can significantly reduce the error in the audible band; however, this improvement does not come without penalty. Oversampling introduces an increase in computational expense, both because the filter must be implemented over more samples, and because the resampling process itself requires computation. Additionally, resampling requires the implementation of anti-aliasing and anti-imaging filters. Performing this filtering with linear-phase, half-band FIR filters, as is common, introduces delay and pre-ringing into the time domain response. The pre-ringing and delay artifacts introduced by the oversampling process are shown in Fig.~\ref{fig:ovrSamp}, which shows the sampled impulse response of the continuous-time prototype filter along with the impulse response of the filter after discretization with the bilinear transform using an oversampling factor of $2$. 

In this paper, all oversampling was performed using linear-phase, half-band FIR interpolators and decimators, created with MATLAB's \textit{designHalfbandFIR} function~\cite{MATLABhalfBand}. The details of the structure of these filters can be viewed in the companion MATLAB live script \cite{playBackEQ}. 

\section{Results \& Discussion}\label{sect:results}

\begin{table*}[!htb] 
\begin{tabular}{lccccccccc}
                                         & \multicolumn{9}{c}{Bark-Weighted RMSE (20Hz-20kHz)}                                                                                                                                                                                                                                                                                                            \\
                                         & \multicolumn{3}{c}{Magnitude (dB)}                                                                                 & \multicolumn{3}{c}{Phase (\degree)}                                                                                      & \multicolumn{3}{c}{Complex}                                                                               \\
{\underline{Method}}                & $\times1$                                    & $\times2$                                   & $\times4$                                     & $\times1$                                   & $\times2$                                   & $\times4$                                      & $\times1$                                    & $\times2$                                     & $\times4$                                     \\ \cline{2-10} 
\multicolumn{1}{l|}{Zero-Order Hold}     & \multicolumn{1}{c}{0.51}            & \multicolumn{1}{c}{0.124}          & \multicolumn{1}{c|}{0.0308}           & \multicolumn{1}{c}{22.5}           & \multicolumn{1}{c}{11.0}           & \multicolumn{1}{c|}{5.47}              & \multicolumn{1}{c}{0.103}           & \multicolumn{1}{c}{0.0498}           & \multicolumn{1}{c|}{0.0246}           \\ \cline{2-10} 
\multicolumn{1}{l|}{Triangle Approx.}    & \multicolumn{1}{c}{1.41}            & \multicolumn{1}{c}{0.261}          & \multicolumn{1}{c|}{0.0620}           & \multicolumn{1}{c}{\textbf{0.509}} & \multicolumn{1}{c}{\textbf{0.015}} & \multicolumn{1}{c|}{\textbf{8.04E-4}} & \multicolumn{1}{c}{0.0180}          & \multicolumn{1}{c}{0.00431}          & \multicolumn{1}{c|}{0.00128}          \\ \cline{2-10} 
\multicolumn{1}{l|}{Impulse Invariant}  & \multicolumn{1}{c}{1.14}            & \multicolumn{1}{c}{0.475}          & \multicolumn{1}{c|}{0.217}            & \multicolumn{1}{c}{21.2}           & \multicolumn{1}{c}{10.8}           & \multicolumn{1}{c|}{5.48}              & \multicolumn{1}{c}{0.145}           & \multicolumn{1}{c}{0.0709}           & \multicolumn{1}{c|}{0.0349}           \\ \cline{2-10}
\multicolumn{1}{l|}{Bilinear Transform}  & \multicolumn{1}{c}{1.39}            & \multicolumn{1}{c}{0.256}          & \multicolumn{1}{c|}{0.0607}           & \multicolumn{1}{c}{0.791}          & \multicolumn{1}{c}{0.187}          & \multicolumn{1}{c|}{0.0463}            & \multicolumn{1}{c}{0.0176}          & \multicolumn{1}{c}{0.00419}          & \multicolumn{1}{c|}{0.00125}          \\ \cline{2-10} 
\multicolumn{1}{l|}{Zero-Pole Matching}  & \multicolumn{1}{c}{0.504}           & \multicolumn{1}{c}{0.123}          & \multicolumn{1}{c|}{0.0305}           & \multicolumn{1}{c}{22.5}           & \multicolumn{1}{c}{11.0}           & \multicolumn{1}{c|}{5.47}              & \multicolumn{1}{c}{0.103}           & \multicolumn{1}{c}{0.0498}           & \multicolumn{1}{c|}{0.0246}           \\ \cline{2-10} 
\multicolumn{1}{l|}{Complex Error Min.}  & \multicolumn{1}{c}{1.34}            & \multicolumn{1}{c}{0.237}          & \multicolumn{1}{c|}{0.0556}           & \multicolumn{1}{c}{1.11}           & \multicolumn{1}{c}{0.259}          & \multicolumn{1}{c|}{0.0644}            & \multicolumn{1}{c}{\textbf{0.0166}} & \multicolumn{1}{c}{\textbf{0.00389}} & \multicolumn{1}{c|}{\textbf{0.00119}} \\ \cline{2-10} 
\multicolumn{1}{l|}{Mag. Error Min.}     & \multicolumn{1}{c}{\textbf{0.0584}} & \multicolumn{1}{c}{\textbf{0.027}} & \multicolumn{1}{c|}{\textbf{0.00802}} & \multicolumn{1}{c}{17.2}           & \multicolumn{1}{c}{8.17}           & \multicolumn{1}{c|}{4.07}              & \multicolumn{1}{c}{0.0716}          & \multicolumn{1}{c}{0.0362}           & \multicolumn{1}{c|}{0.0183}           \\ \cline{2-10} 
\multicolumn{1}{l|}{Nyquist Band Trans.} & \multicolumn{1}{c}{0.430}           & \multicolumn{1}{c}{0.132}          & \multicolumn{1}{c|}{0.0321}           & \multicolumn{1}{c}{15.1}           & \multicolumn{1}{c}{6.99}           & \multicolumn{1}{c|}{3.46}              & \multicolumn{1}{c}{0.123}           & \multicolumn{1}{c}{0.0448}           & \multicolumn{1}{c|}{0.0174}           \\ \cline{2-10} 
\end{tabular}
\caption{Magnitude, phase, and complex Bark-weighted mean squared error between the sampled continuous-time prototype filter and the discrete-time filters generated using each method. Each measurement was made at \SI{1}{\hertz} resolution over the bandwidth \SI{20}{\hertz}-\SI{20}{\kilo\hertz}. The base sample rate in each case was \SI{48}{\kilo\hertz}. The factors above each column indicate the amount of oversampling that was implemented. $\times1$ indicates no oversampling, $\times2$ indicates a oversampling factor of $2$ for a maximum sampling frequency of \SI{96}{\kilo\hertz}, and $\times4$ indicates a oversampling factor of $4$ for a maximum sampling frequency of \SI{192}{\kilo\hertz}. The lowest error values in each column are bolded.}
\label{tbl:accuracy}
\end{table*}

\begin{figure*}[h]
\begin{center}
\includegraphics[width=0.95\textwidth]{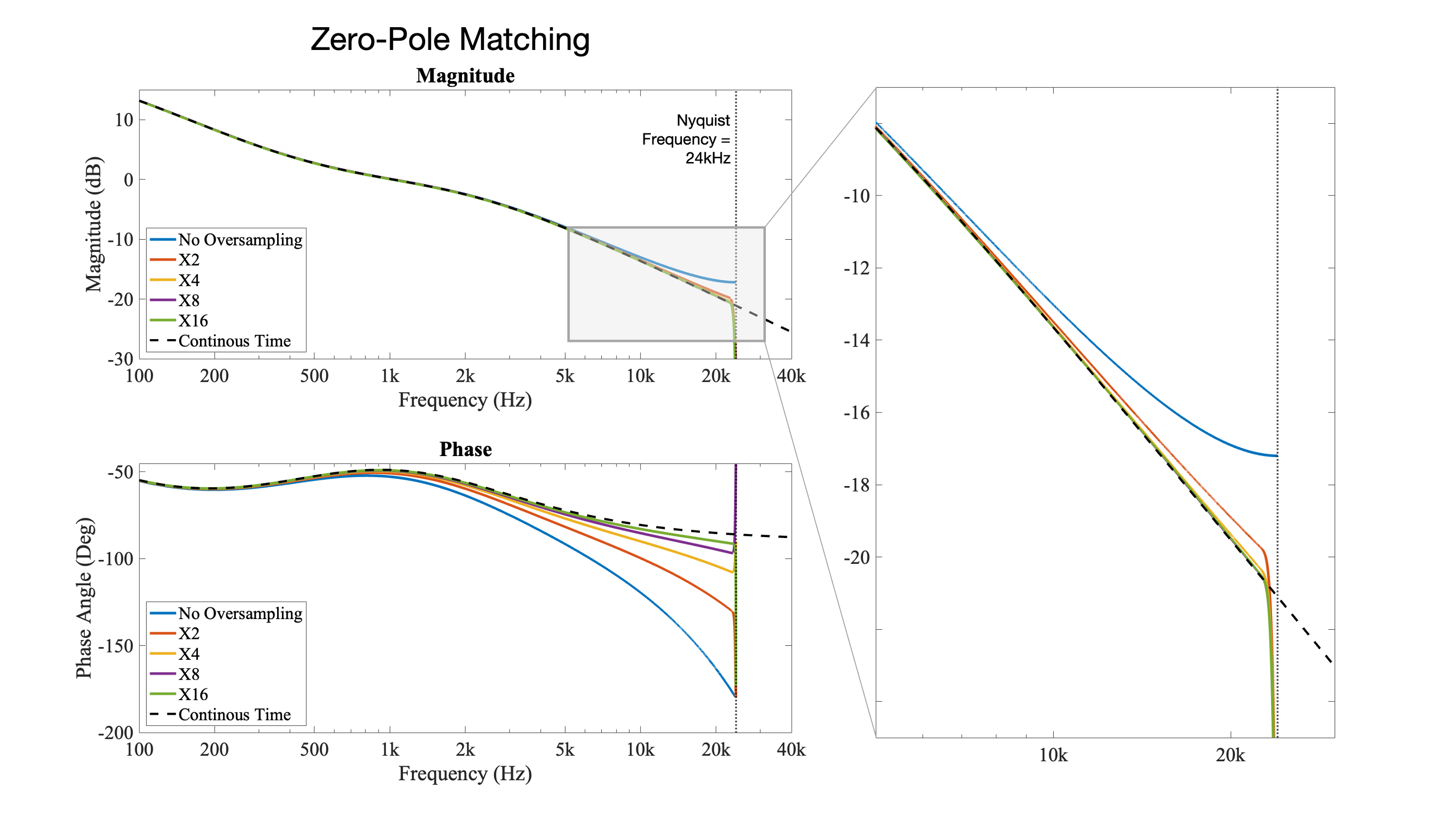}
\caption{Magnitude and phase responses of the discrete-time filters generated using the Zero-Pole Matching method with differing amounts of oversampling, along with the sampled magnitude and phase response of the continuous-time prototype filter. All discrete filters have a base sample rate of \SI{48}{\kilo\hertz}.}
\label{fig:ZPM}
\end{center}
\end{figure*}

\subsection{Error Weighting} \label{sect:weighting}
A method for weighting error in the frequency domain was implemented such that perceptually equivalent~deviations in the responses of the filters account for equivalent increases in the overall error reported. This was accomplished by assigning a weight to each frequency bin that is the inverse of the bandwidth of the Bark band in which that frequency falls. Bark, or critical, bands represent, "a subdivision more closely related to the manner in which the ear itself appears to carry out the process," compared to purely linear or purely logarithmic groupings~\cite{zwicker1961subdivision}. This weighting scheme has the result that each Bark band has equal weight in the error calculations. A similar weighting system is recommended in \cite{darabundit2022nyquist} for use in the optimization of a free parameter that influences frequency warping. In this work, this weighting scheme is used in the evaluation of the magnitude and phase error in the results section, as well as in the minimizations required by the Complex Error Minimization and Magnitude Error Minimization methods. Additionally, this weighting is used in the optimization of the free parameter, $\gamma$, used in the Nyquist Band Transform method~(see Appendix~\ref{NBT}).

\subsection{Error Metrics} \label{sect:errorMetrics}

The prototype RIAA filter was discretized using each method described in Sect. \ref{sect:methods}. A base sampling rate of \SI{48}{\kilo\hertz} was chosen because the Nyquist frequency at this sample rate is close enough to \SI{20}{\kilo\hertz} for the error in the audible band to be pronounced. The frequency-domain performance of each discretization is characterized by three metrics. The accuracy of the magnitude response is calculated as,

\begin{equation}\label{eq:RMSEmag}  
    RMSE_{mag} = \sqrt{\frac{\sum_{k=0}^{L-1}w_k\left(20\log_{10}\left[\frac{|H(j\omega_k)|}{|\hat{H}(j\omega_k)|}\right]\right)^{2}}{\sum_{k=0}^{L-1}{w_k}}},
\end{equation}

where $L$ is the number of frequency bins to be analyzed, ${H}$ is the sampled complex frequency response of the continuous-time prototype filter, $\hat{H}$ is the complex frequency response of the discretized filter, $\omega$ is a length-$n$ vector of angular frequencies, $w$ is a length-$L$ array of Bark weights, and $k$ is an indexing variable. To accommodate a \SI{1}{\hertz} resolution over the \SI{20}{\hertz}-\SI{20}{\kilo\hertz} bandwidth, $L=19,981$ is chosen. 

The phase response error\footnote{Note that when reporting the phase response error for the oversampled tests, a linear phase offset was added to the discrete-time filter phase response to compensate for the group delay introduced by the FIR anti-aliasing and anti-imaging filters.} is calculated as,
\begin{equation}\label{eq:RMSEang}  
    RMSE_{ang} = \sqrt{\frac{\sum_{k=0}^{L-1}w_k\left(\angle H(j\omega_k)-\angle\hat{H}(j\omega_k)\right)^{2}}{\sum_{k=0}^{L-1}{w_k}}}.
\end{equation}
The complex response error is calculated as,
\begin{equation}\label{eq:RMSEcomp}  
    RMSE_{comp} = \sqrt{\frac{\sum_{k=0}^{L-1}w_k|H(j\omega_k)-\hat{H}(j\omega_k)|^{2}}{\sum_{k=0}^{L-1}{w_k}}}.
\end{equation}
This metric equates to the square root of the mean of the squared Euclidean distance between the points in the complex plane.

\subsection{Accuracy}
The accuracy with which each discretization method reproduces the continuous-time response of the prototype filter as measured by the metrics above is presented in Table \ref{tbl:accuracy} for oversampling factors of $\times2$ and $\times4$ as well with no oversampling. The lowest magnitude, phase, and complex errors were achieved by the Magnitude Error Minimization, Triangle Approximation, and Complex Error Minimization methods, respectively. The first and third results are unsurprising as these methods iteratively minimize an error metric that is related to the ones that we are using for evaluation. While these methods are highly accurate, as discussed below, they are the most computationally expensive of the methods, and are therefore potentially unsuitable for some real-time applications. Of the non-iterative methods, the NBT exhibits the lowest magnitude error with no oversampling, and Zero-Pole Matching achieves the lowest error when oversampling is implemented. In terms of complex accuracy, of the non-iterative methods, the Bilinear Transform exhibits the lowest error. For each method, all three metrics are improved as the oversampling factor is increased.  

The sampled response of the continuous-time prototype filter along with the responses of discrete-time filters generated using the Zero-Pole matching method at several levels of oversampling are shown in Fig.~\ref{fig:ZPM}. From this figure, we can see that the responses are quite accurate in the lower band, and begin to fail as we approach the Nyquist frequency. There is also an improvement in accuracy as the amount of oversampling is increased. The effect of the anti-aliasing filters for the oversampled cases can be observed in the steep dropoff in the magnitude response at the Nyquist frequency. Although only the Zero-Pole matching method is shown, these general trends are repeated for all methods. Fig.~\ref{fig:ZPM} was generated with the companion MATLAB live script, which can be used to make similar figures for any of the presented methods \cite{playBackEQ}.

\subsection{Computational Efficiency}
When a cutting curve is unknown or is in question, it is common that a restoration engineer will need to audition multiple playback equalizations. Under these conditions, it is beneficial for the operator to be able to adjust the filter parameters in real time during playback. For this reason, along with the accuracy with which each method reproduces the continuous-time response, we must also consider the computational efficiency of the algorithm. Quantifying the computational cost of each method is beyond the scope of this article, but we can say that, in general, the iterative approaches are likely to be the slowest, and methods that entail a simple substitution for $s$ in terms of $z$, like the Bilinear Transform and Zero-Pole Matching methods, are likely to be the fastest. The companion MATLAB live script \cite{playBackEQ} includes a measurement of the time required to discretize the filters using each method for the system on which the script is running, which can provide a rough idea of the relative computational efficiencies of each approach.

\section{Summary}
In this article, we have surveyed eight methods for discretizing continuous-time systems, and explored each of them in the context of performing phonograph playback equalization in the digital domain. The results showed that the iterative methods with maximum oversampling provided the most faithful reproduction of the continuous-time magnitude and complex responses, but that methods with lower computational overhead can still generate very accurate filters. The intention of this article is not to determine which of these methods is best, but to quantify the compromises that each one entails in a way that might allow someone developing a system for playback equalization to make informed choices for their specific application and platform.

Each method presented here can be explored in detail using the interactive companion MATLAB live script found at \url{https://doi.org/10.60593/ur.d.26503432} \cite{playBackEQ}.

\bibliographystyle{jaes}

\bibliography{refs}

\appendix{}
\section*{APPENDIX}
\section{Nyquist Band Transform Algorithm}\label{NBT}
The NBT is accomplished in four steps. In the first transform, the region $\omega = [0,\omega_0)$ is mapped back to itself, and also flipped and stretched to span the region $\omega = [\omega_0,\infty)$. In this mapping, dc is mapped to dc and $\infty$, and $\omega_0$ is mapped to itself \footnote{In the original work~\cite{darabundit2022nyquist}, a distinct truncation frequency, $\omega_c$, and a distinct critical frequency, $\omega_0$, are specified. For our purposes, we will set $\omega_c = \omega_0 = \pi f_{s}$ rad/s, where $f_{s}$ is the desired sampling frequency in Hz.}. This transform takes the form,
\begin{equation}\label{eq:firstTransform}   
     G_1(s) = \frac{-2\omega_0^{2}s}{s^{2}-\omega_0^{2}}.
\end{equation}
For our prototype filter from Eq. \ref{eq:H}, the first stage is,
\begin{equation}\label{eq:firstTransformMatrix}  
    \textbf{A}_1\textbf{c}=\textbf{k},
\end{equation}
where $\textbf{A}_1$ is the first transform in matrix form,
\begin{equation}\label{eq:firstTransformMatrix}  
    \begin{bmatrix}
    0               & 0                 & 1\\
    0               & -2\omega_0^{2}    & 0\\
    4\omega_0^{4}   & 0                 & -2\omega_0^{2}\\
    0               & 2\omega_0^{4}     & 0\\
    0               & 0                 & \omega_0^{4} \\
    \end{bmatrix},
\end{equation}
\textbf{c} is a vector of the numerator or denominator coefficients of the prototype continuous-time filter,
\begin{equation}\label{eq:cMatrix}      
    \begin{bmatrix}
    c_2 \\
    c_1 \\
    c_0 \\
    \end{bmatrix},
\end{equation}
and $\textbf{k}$ is a vector of the numerator or denominator coefficients of the resulting filter,
\begin{equation}\label{eq:kMatrix}      
    \begin{bmatrix}
    k_4 \\
    k_3 \\
    k_2 \\
    k_1 \\
    k_0 \\
    \end{bmatrix}.
\end{equation}

All stages of the NBT prior to the fourth stage, the bilinear transform, are carried out independently for both the numerator and the denominator of the transfer function. The subscripts of the coefficients indicate the power of $s$ in the term they are associated with; i.e., $k_4$ is the coefficient for the $s^4$ term, etc. For the numerator of our prototype filter, an $s^2$ term, $c_2=0$, has been added so that the number of terms in the numerator is the same as the number of terms in the denominator. Notice that the output of this transform is a filter with numerator and denominator orders of $2N$, where $N$ is the order of the prototype filter.

The filter that results from the first transform is neither stable nor minimum phase. To address this, in the second stage of the NBT, any poles and zeros that fall on the right side of the $s$-plane are reflected over the $j\omega$ axis, resulting in a stable, minimum phase filter. This process generates a new vector of numerator or denominator coefficients, $\textbf{m}$, with the same form as Eq.~\ref{eq:kMatrix}.

In the third stage of the NBT, the region $\omega = [0,\omega_0)$ is mapped to $\omega = [0,\infty)$. This is accomplished by the \textit{inverse} second transform. The \textit{forward} second transform is defined as,
\begin{equation}\label{eq:firstTransform}   
     G_2(s) = \frac{\gamma s}{s^{2}+\omega_0^{2}},
\end{equation}
where $\gamma$ is a free parameter that affects frequency warping. The value of $\gamma$ can be tuned to perfectly map the continuous-time response at a single frequency to that same frequency in discrete time in the same way that pre-warping affects the result of a bilinear transform. As with the bilinear transform, this can be useful if there is a specific frequency of interest. Alternatively, $\gamma$ can be set to reduce frequency warping across the spectrum. In this paper, $\gamma$ was determined by minimizing the Bark-weighted mean squared error~(see Sect.~\ref{sect:results}) between the original frequencies and the frequencies as mapped by the NBT from \SI{20}{\hertz} to \SI{20}{\kilo\hertz}. Values of $\gamma$ for each desired sample rate were determined this way.

For our prototype filter, the second forward transform in matrix form, $\textbf{A}_2$, is,
\begin{equation}\label{eq:secondTransformMatrix}   
    \begin{bmatrix}
    0               & 0                     & 1\\
    0               & \gamma                & 0\\
    \gamma^{2}      & 0                     & 2\omega_0^{2}\\
    0               & \omega_0^{2}\gamma    & 0\\
    0               & 0                     & \omega_0^{4} \\
    \end{bmatrix}.
\end{equation}
To obtain the inverse second transform, the Moore-Penrose inverse of $\textbf{A}_2$ is computed to generate $\textbf{A}_{2}^{+}$. The third stage is then,
\begin{equation}\label{eq:firstTransformMatrix}  
    \textbf{A}_2^{+}\textbf{m}=\textbf{p},
\end{equation}
where $\textbf{p}$ is a new vector of numerator or denominator coefficients with the same form as Eq. \ref{eq:cMatrix}.

In the final stage, a standard bilinear transform is performed on the pre-mapped continuous-time filter to obtain the discretized filter.

\end{document}